\documentclass[showpacs]{revtex4}
\usepackage{epsfig}
\def\beq{\begin{equation}}
\def\eeq{\end{equation}}
\def\enq{\end{equation}}
\def\beqa{\begin{eqnarray}}
\def\eeqa{\end{eqnarray}}
\def\bnqa{\begin{eqnarray}}
\def\enqa{\end{eqnarray}}

\def\GeV{\nobreak\,\mbox{GeV}}

\def\pli{p^\prime}
\def\ql{{p^\prime}^2}
\def\mli{{M^\prime}^2}

\def\BS{B_{s0}}
\newcommand{\rag}{\rangle}
\newcommand{\lag}{\langle}
\def\lb{\label}
\def\nn{\nonumber}
\def\qq{\lag\bar{q}q\rag}
\def\ss{\lag\bar{s}s\rag}
\def\mix{\lag\bar{q}g\sigma.Gq\rag}
\def\mixs{\lag\bar{s}g\sigma.Gs\rag}
\def\Gd{\lag g^2G^2\rag}
\def\G3{\lag g^3G^3\rag}
\def\La{\Lambda}
\def\al{\alpha}

\begin{document}
\title{The $B_{s0}$ meson and the $B_{s0}B K $ coupling from QCD sum rules}
\author{M.E. Bracco}
\affiliation{Faculdade de Tecnologia, Universidade do Estado do Rio de 
Janeiro, 
Rod. Presidente Dutra Km 298, Polo Industrial, CEP , Resende, RJ, Brazil}
\author{ M. Nielsen}
\affiliation{Instituto de F\'{\i}sica, Universidade de S\~{a}o Paulo, 
C.P. 66318, 05389-970 S\~{a}o Paulo, SP, Brazil.}

\begin{abstract}
We evaluate the mass of the $B_{s0}$ scalar meson and the coupling constant in  
the $B_{s0} B K$ vertex in the framework of QCD sum rules. We consider  the
$B_{s0}$ as a tetraquark state to evaluate its mass. We get $m_{B_{s0}}=
(5.85\pm
0.13)~\GeV$, which is in agreement, considering the uncertainties, with 
predictions supposing it as a $b\bar{s}$ state
or a $B\bar{K}$ bound state with $J^{P}=0^+$. To evaluate the $g_{B_{s0}B K}$ 
coupling we use the three point correlation functions of the vertex, 
considering $ B_{s0} $ as a normal $b\bar{s}$ state. The obtained coupling 
constant is: $g_{B_{s0} B K} =(16.3 \pm 3.2)~\GeV$. This number  is in agreement
with light-cone QCD sum rules calculation. We have also compared the decay width
of the $\BS\to BK$ process considering the $\BS$ to be a $b\bar{s}$ state
and a $BK$ molecular state. The width obtained for the $BK$ molecular state
is twice as big as the width obtained for the $b\bar{s}$ state. Therefore, we 
conclude that with the knowledge of the mass and the decay width of the
$\BS$ meson, one can discriminate between the different theoretical proposals
for its structure.
\end{abstract}


\maketitle

\section{Introduction}

In recent years the observation of heavy flavor hadrons have raised strong 
interest in interpreting these states as multiquark states, like tetraquarks
states and hadronic molecules. Famous examples are the scalar $D_{s0}(2317)$
and the axial $D_{s1}(2460)$ charmed mesons, and the $X(3872)$ hidden charm 
meson. In the bottom sector, the recent observations of the $J^P=1^+$
$B_{s1}(5830)$ by the CDF collaboration \cite{cdf} and the $J^P=2^+$
$B_{s2}(5840)$ by the CDF  and D0 collaborations \cite{cdf,d0} enrich the spectrum
of the bottom-strange system and estimulate our interest in the possible
 interpretation of these states as multiquark states. In particular the yet
unobserved $J^P=0^+$ $B_{s0}$ state, could be very broad and, therefore, very 
difficult to be observed, if it is a multiquark system with mass above the $BK$ 
threshold. There are already some predictions for the $B_{s0}$ mass supposing
it is a $B\bar{K}$ bound state \cite{gscpz}, as a $b\bar{s}$ state \cite{wang},
and as a mixture between a $b\bar{s}$ and a $(bq)(\bar{s}\bar{q})$ states
\cite{vvf}. 
Although the structure for the $B_{s0}$ in these calculations is very 
different, the predictions for its mass are very similar: $(5.725\pm0.039)~\GeV$
in ref.~\cite{gscpz}, $(5.70\pm0.11)~\GeV$ in ref.~\cite{wang} and $5.68 ~\GeV$
in ref.~\cite{vvf}, for a state with 30\% of the four-quark component.

There are also predictions for the $B_{s0}BK$ coupling constant, supposing
the $B_{s0}$ to be a $B\bar{K}$ bound state \cite{gscpz} and a $b\bar{s}$ state 
\cite{wang2}, and for the $B_{s0}B_s\pi$ 
coupling constant, supposing the $B_{s0}$ to be a $B\bar{K}$ bound state 
\cite{fglm}. The knowledge of the coupling constant at the  $B_{s0}BK$ vertex is 
very important since the decay width in this channel, the most important
channel if it is allowed, can give an idea of the total width of 
the state. For one still unobserved state, it is very important for the 
experimentalists to have theoretical predictions not only about its mass, but 
also about its width. As an example, the mass of the scalar $D_{s0}(2317)$
meson is below the $DK$ threshold, therefore the decay $D_{s0}(2317)\to DK$ is
not allowed. As a consequence, the $D_{s0}(2317)$ is very narrow since the main 
decay channel for this state is the isospin violating mode $D_{s0}(2317)\to D\pi$.
The same could happen with the scalar $B_{s0}$ if its mass is bellow the
$BK$ threshold. By the other hand, if its mass is above the $BK$ threshold,
two scenarios are possible: 
\begin{itemize}
\item{ it can be a very broad state if it is a tetraquark state, like the light 
scalars, since the decay $B_{s0}\to BK$ will be super allowed}

\item{ or it can still be narrow if it is a normal $b\bar{s}$ state, like the
new recently observed states $B_{s1}(5830)$ and $B_{s2}(5840)$, if the
coupling at the vertex $B_{s0}BK$ is not very large.}
\end{itemize}
In the last case, the knowledge of the $B_{s0}BK$ coupling constant is indeed
very important, and could help the experimentalists in the search for such state.

In this work we use the QCD sum rule (QCDSR) approach 
\cite{svz,rry,SNB} to study the mass of the $B_{s0}$ 
supposing it is a tetraquark state in a diquark-antidiquark
configuration, similar to the supposition made for the $D_{s0}(2317)$
scalar meson in ref.~\cite{pec}. We get a bigger mass as compared with 
predicitions supposing it as a $b\bar{s}$ state \cite{wang}
or a $B\bar{K}$ bound state \cite{gscpz} but, considering the 
uncertainties, still compatible with these predictions. Since the mass obtained 
is above the $BK$ threshold, as explained above, a tetraquark scalar $B_{s0}$
will be very broad and, therefore, very difficult to be observed. 

Since the prediction for the mass of $B_{s0}$, supposing it is a normal $b\bar{s}$
state: $(5.70\pm0.11)~\GeV$ \cite{wang}, can be above the $BK(5774)$ threshold, as
explained above, it is very important to know the $B_{s0}BK$ coupling constant,
therefore, in this work we also evaluate the
$B_{s0}BK$ coupling constant supposing the $B_{s0}$ to be a normal $b\bar{s}$ 
state.

This paper is organized as follows: in Sec. II we introduce the two-point function
to evaluate the mass of the tetraquark state. In Sec. III  we study
the three-point function for the $B_{s0}BK$ vertex. In Sec. IV we present our 
conclusions.

\section{Sum rule for the  mass of the $ B_{s0}$ scalar meson}

In the QCDSR approach, the short range perturbative QCD is
extended by an OPE expansion of the correlator, which results in 
a series in powers of
the squared momentum with Wilson coefficients. The convergence at low
momentum is improved by using a Borel transform. The expansion involves
universal quark and gluon condensates. The quark-based calculation of
a given correlator is equated to the same correlator, calculated using
hadronic degrees of freedom via a dispersion relation, providing sum rules
from which a hadronic quantity can be estimated. 

Considering the $B_{s0}$ scalar meson as a $S$-wave bound state of a 
diquark-antidiquark pair, and considering the diquark in  a spin zero 
colour anti-triplet, a possible current describing such state is given by:

\beq
j={\epsilon_{abc}\epsilon_{dec}\over\sqrt{2}}\left[(u_a^TC
\gamma_5b_b)(\bar{u}_d\gamma_5C\bar{s}_e^T)+(d_a^TC
\gamma_5b_b)(\bar{d}_d\gamma_5C\bar{s}_e^T)\right],
\label{int}
\enq
where $a,~b,~c,~...$ are colour indices and $C$ is the charge conjugation
matrix.

The QCDSR for the bottom-strange scalar meson is constructed from the two-point
correlation function
\beq
\Pi(q)=i\int d^4x ~e^{iq.x}\lag 0 |T[j(x)j^\dagger(0)]|0\rag.
\lb{2po}
\enq

In the OPE side we work at leading order in $\alpha_S$ and consider condensates 
up to dimension eight. We treat the strange quark as a light quark and consider
the diagrams up to order $m_s$. 
In ref.~\cite{jido} it was shown that the $q\bar{q}$ annihilation diagrams
are more important for the 4-quark correlators than for the normal 2-quark
mesonic correlators. Therefore, the  $q\bar{q}$ annihilation diagrams can not
be neglected a priori. Also, due to these $q\bar{q}$ annihilation diagrams, 
the current in Eq.~(\ref{int}) can mix with a  two-quark $b\bar{s}$ current. 
Mixed tetraquark two-quark currents were considered to the study of the light 
scalars \cite{oka24} and also for the study of the $X(3872)$ meson 
\cite{x24,report} in the framework of QCDSR. In this work we will not consider 
the $q\bar{q}$ annihilation diagrams neither the mixed tetraquark two-quark 
current. Therefore, our calculation will provide only the 4-quark 
contributions to the  4-quark correlator in Eq.~(\ref{2po}).

The correlation function in the OPE side can be written in terms of a 
dispersion relation:
\beq
\Pi^{OPE}(q^2)=\int_{m_b^2}^\infty ds {\rho(s)\over s-q^2}\;,
\lb{ope}
\enq
where the spectral density is given by the imaginary part of the correlation
function: $\rho(s)={1\over\pi}\mbox{Im}[\Pi^{OPE}(s)]$.

In the phenomenological side
the coupling of the scalar $B_{s0}$ meson to the scalar current in Eq.~(\ref{int})
  can be parametrized in terms of a parameter $\lambda$ as:
$\lag 0 | j|B_{s0}\rag =\lambda$. Therefore, the
phenomenological side of Eq.~(\ref{2po}) can be written in terms of $\lambda$ as:
\beq
\Pi^{phen}(q^2)={\lambda^2\over m_{\BS}^2-q^2}+\cdots\;,
\lb{phe}
\enq
where the dots denote higher resonance contributions that will be 
parametrized, as usual, through the introduction of the continuum threshold
parameter $s_0$ \cite{io1}.

 After making a Borel
transform on both sides, and transferring the continuum contribution to
the OPE side, the sum rule for the scalar meson $\BS$ can be written as
\beq
\lambda^2e^{-m_{\BS}^2/M^2}=\int_{m_b^2}^{s_0}ds~ e^{-s/M^2}~\rho(s)\;,
\lb{sr}
\enq
where $\rho(s)=\rho^{pert}(s)+\rho^{m_s}(s)+\rho^{\qq}(s)+\rho^{\lag G^2\rag}
(s)+\rho^{mix}(s)+\rho^{\qq^2}(s)+\rho^{\lag G^3\rag}(s)$, with
\beq
\rho^{pert}(s)={1\over 2^{10} 3\pi^6}\int_\La^1 d\al\left({1-\al\over\al}
\right)^3(m_b^2-s\al)^4,
\enq
\beq
\rho^{m_s}(s)=0
\enq
\beq
\rho^{\qq}(s)={1\over 2^{6}\pi^4}\int_\La^1 d\al~{1-\al\over\al}
(m_b^2-s\al)^2\bigg[
-\qq\left(2m_s+m_b{1-\al\over\al}\right)+m_s\ss\bigg],
\enq
\beq
\rho^{\lag G^2\rag}(s)={\Gd\over 2^{10}\pi^6}\int_\La^1 d\al~(m_b^2-s\al)
\left[{m_b^2\over9}\left({1-\al\over\al}\right)^3+
(m_b^2-s\al)\left({1-\al\over2\al}+{(1-\al)^2\over4\al^2}\right)
\right],
\enq
\beq
\rho^{\lag G^3\rag}(s)={\G3\over 2^{12} 9\pi^6}\int_\La^1 d\al\left({1-\al
\over\al}\right)^3(3m_b^2-s\al),
\enq
\beqa
\rho^{mix}(s)&=&{1\over 2^{6}\pi^4}\int_\La^1 d\al~(m_b^2-s\al)\bigg[-{m_s\mixs
\over6}+\mix\bigg(-m_s(1-\ln(1-\al))
\nn\\
&-&m_b{1-\al\over\al}\left(1-{1-\al\over2\al}\right)\bigg)
\bigg]
\enqa
\beq
\rho^{\qq^2}(s)=-{\qq\over 24\pi^2}\left[\ss\left(2m_b^2-s-{m_b^4\over s}
\right)+m_sm_b\left(\ss-2\qq\right)\left(1-{m_b^2\over s}\right)\right].
\enq
The lower limit of the integrations is given by $\La=m_b^2/s$.

\subsection{Results for the mass}

In the numerical analysis of the sum rules, the values used for the quark
masses and condensates are \cite{SNB,SNCB,narpdg,SNG}: 
$m_s=0.13\,\GeV$, $m_b(m_b)=(4.24\pm 0.06)\,\GeV$, 
$\lag\bar{q}q\rag=\,-(0.23)^3\,\GeV^3$,
$\langle\overline{s}s\rangle\,=0.8\lag\bar{q}q\rag$,  
$\lag\bar{q}g\sigma.Gq\rag=m_0^2\lag\bar{q}q\rag$, $\lag\bar{s}g\sigma.Gs\rag=
m_0^2\lag\bar{s}s\rag$ with $m_0^2=0.8\,\GeV^2$, $\lag g^2G^2\rag=0.88~\GeV^4$
and $\lag g^3G^3\rag=0.045~\GeV^6$. 

\begin{figure}[h]
\centerline{\epsfig{figure=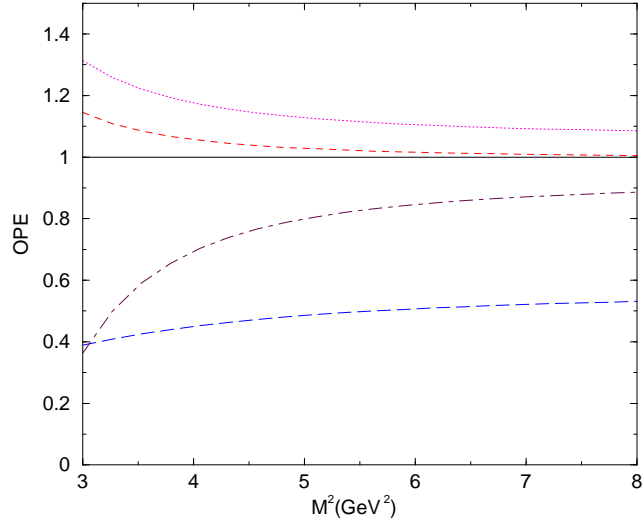,height=70mm}}
\caption{The OPE convergence for the $J^{P}=0^{+},~B_{s0}$ 
meson in the region
$3 \leq M^2 \leq8~\GeV^2$ for $\sqrt{s_0} = 6.6$ GeV.  We plot the 
relative contributions starting with the perturbative contribution 
(long-dashed line), and each other line represents the 
relative contribution after adding of one extra condensate in the expansion: 
+ quark condensate  (dashed line), 
+ gluon condensate (dotted line), + mixed condensate
(dot-dashed line), + four-quark condensate (solid line).}
\label{figconv}
\end{figure}

The Borel window is determined by analysing the OPE convergence, the Borel
stability  and the pole contribution. To determine the minimum value of the 
Borel mass we impose that the contribution of the higher dimension condensate 
should be smaller than 20\% of the total contribution: $M^2_{min}$ is such that
\beq
\left|{\mbox{OPE summed up dim n-1 }(M^2_{min})\over\mbox{total 
contribution }(M^2_{min})}\right|=0.8.
\label{mmin}
\enq

In Fig.~\ref{figconv} we show the contribution of all the terms in the
OPE side of the sum rule. From this figure we see that only  for $M^2\geq 5.3$ 
GeV$^2$ the contribution of the dimension-6 condensate is around 20\% of the
total contribution. However, for such large value of the Borel mass there is
no dominance of the pole contribution for  $\sqrt{s_0} = 6.6$ GeV. We interpret
this as an indication that the dimension-6 condensate does not saturate the OPE.
To improve the OPE convergence, we also include the dimension-8 condensate:

\beqa
\rho^{D=8}(s)&=&{m_0^2\qq\over 24\pi^2}\bigg[\ss+{m_sm_b\over4}\bigg((4\qq
-ss)\int_0^1{d\alpha\over1-\alpha}\delta\left(s-{m_b^2\over1-\alpha}\right)
-(4\qq-2\ss)\delta(s-m_b^2)\bigg)\bigg]
\enqa

\begin{figure}[h]
\centerline{\epsfig{figure=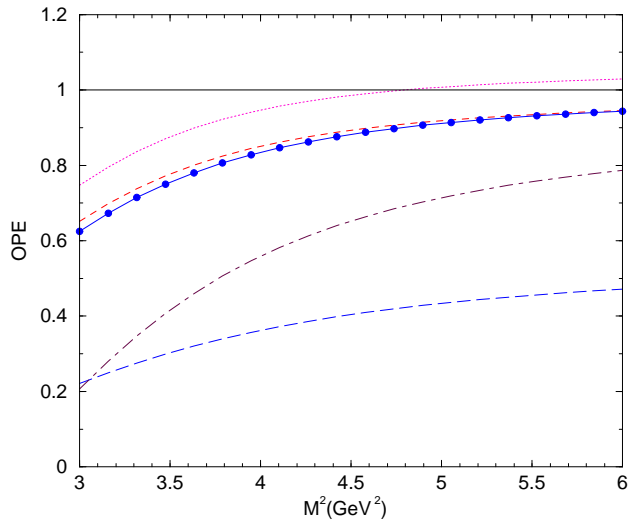,height=70mm}}
\caption{Same as Fig.~1 for $\sqrt{s_0} = 6.7$ GeV:
perturbative contribution (long-dashed line), 
+ quark condensate  (dashed line), 
+ gluon condensate (dotted line), + mixed condensate
(dot-dashed line), + four-quark condensate (solid line with dots)+ dimension-8 
condensate (solid line).}
\label{figconv2}
\end{figure}

From Fig.~\ref{figconv2} we see that for $M^2\geq 4.0$ 
GeV$^2$ the contribution of the dimension-8 condensate is smaller than 20\% 
of the
total contribution. Therefore, the inclusion of the  dimension-8 condensate
improved the OPE convergence of the sum rule. We  fix the lower
value of $M^2$ in the sum rule window as $M^2_{min}= 4.0$ GeV$^2$.
One should note that a complete evaluation of higher dimension condensates 
contributions require more involved analysis including a non-trivial choice
of the factorization assumption basis \cite{BAGAN}. Therefore, in this work
we do not consider condensates with dimension higher than 8.

The maximum value of the Borel mass is 
determined by imposing that the pole contribution must be bigger than the 
continuum contribution: $M^2_{max}$ is such that
\beq
{\int_{m_b^2}^{s_0}ds~e^{-s/M^2_{max}}\rho(s)\over
\int_{m_b^2}^\infty ds~e^{-s/M^2_{max}}\rho(s)}= 0.5.
\label{mmax}
\enq

From a physical point of view, the continuum threshold parameter 
is related with the value of the mass of the first excited state, that has
 the same quantum numbers of the studied state. In general, the 
mass of the first excited state state is around 0.5 GeV above the mass
of the low-lying state. Therefore, the continuum threshold can be 
related with the mass of the low-lying state, $H$, through the relation: 
$s_0\sim(m_H+0.5~\GeV)^2$. To choose a good  range of the values of $s_0$ 
we extract the mass from the sum rule, for a given $s_0$, and accept such 
value if the obtained mass is in the range 0.4 GeV to 0.6 GeV smaller 
than $\sqrt{s_0}$. Using this criterion,
we obtain $s_0$ in the range $6.4\leq \sqrt{s_0} \leq6.7$ GeV. However,
for $\sqrt{s_0}= 6.4~\GeV$ the allowed Borel region is very small, therefore, 
we only consider values of $\sqrt{s_0}$ in the range $6.5\leq \sqrt{s_0}\leq
6.7~\GeV$. We show in Table \ref{tablebo} the values of 
$M_{max}$ for different values of $\sqrt{s_0}$.

\begin{table}[h]
\begin{tabular}{|c|c|}  \hline
$\sqrt{s_0}~(\GeV)$ & $M^2_{max}(\GeV^2)$  \\
\hline
 6.5 & 4.30 \\
\hline
 6.6 & 4.47 \\
\hline
6.7 & 4.66 \\
\hline
\end{tabular}
\caption{Upper limits in the Borel window for the $0^+,~
B_{s0}$ meson
obtained from the sum rule for different values of $\sqrt{s_0}$.}
\label{tablebo}
\end{table}

As pointed out in ref.~\cite{jido}, the determination of a sufficiently wide
Borel window is the most important step for the application of the sum rule.
In particular, without imposing correct criteria on the determination
of the Borel window, artefacts as the appearance of pseudopeaks \cite{jido},
could spoil the validity of the QCDSR results. As explained above we have used
the two most important criteria to the determination of the Borel window:
OPE convergence in Eq.~(\ref{mmin}) and pole  contribution 
dominance in Eq.~(\ref{mmax}). Therefore, we do believe that the QCDSR
studied here can be used to extract physical information about the
$B_{s0}$ scalar meson. 

The resonance mass, $m_{\BS}$, can be obtained by taking the derivative
of Eq.~(\ref{sr}) with respect to $1/M^2$ and dividing it by
Eq.~(\ref{sr}):
\beq
m_{\BS}^2={\int_{m_b^2}^{s_0}ds ~e^{-s/M^2}~s~\rho(s)\over\int_{m_b^2}^{s_0}ds 
~e^{-s/M^2}~\rho(s)}\;.
\lb{m2}
\enq

\begin{figure}[h]
\centerline{\epsfig{figure=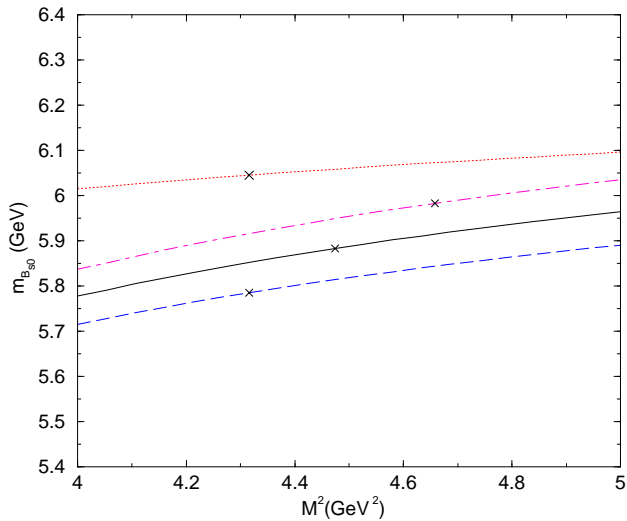,height=70mm}}
\caption{The $\BS$ meson mass, described with a diquark-antidiquark 
current, as a function of the sum rule parameter
($M^2$) for $\sqrt{s_0} =6.5$ GeV (dashed line), $\sqrt{s_0} =6.6$ GeV (solid 
line) and $\sqrt{s_0} =6.7$ GeV (dot-dashed line). The dotted line shows
the result obtained for the mass, for $\sqrt{s_0} =6.6$ GeV, considering only 
the condensates up to
dimension 6. The crosses indicate the upper  limits in the Borel region.}
\label{figm}
\end{figure}

In  Fig.~\ref{figm} we show the obtained mass as a function of the Borel mass,
for different values of $\sqrt{s_0}$.
From this figure we see that the Borel stability is good in the allowed Borel
 window for all considered values of $s_0$. This is different from the case of
the scalar charmed-strange meson $D_{s0}(2317)$, where no Borel window could
be determined using the above mentioned criterious \cite{tetrarev}.
For completeness we also include in this figure (through the dotted line)
the result obtained for the mass when considering only condensates up to 
dimension 6. We see that the inclusion of the dimension-8 condensates not only
improves the OPE convergence but also reduces the mass of the state that couples
with the current in Eq.~(\ref{int}).

Considering the variations on the quark masses, the quark condensate and on 
the continuum 
threshold discussed above, in the Borel window considered here our results 
for the ressonance mass is:
\beq
m_{\BS}=(5.85\pm0.13)~\GeV,
\label{mbs0}
\enq
which is compatible, considering the uncertainties, with the predictions from 
ref.~\cite{gscpz}: $(5.725\pm0.039)~\GeV$, where the $\BS$ is considered as 
a $B\bar{K}$ bound state, and from ref.~\cite{wang}: $(5.70\pm0.11)~\GeV$, 
where the  $\BS$ is considered as a normal $b\bar{s}$ state. 

Since we have not considered the $q\bar{q}$ annihilation diagrams in our 
calculation, the result in Eq.~(\ref{mbs0}) gives only the 4-quark contribution 
to the mass of the $B_{s0}$. Therefore, the result in Eq.~(\ref{mbs0}) is also in 
agreement with the findings in ref.~\cite{vvf}, where the authors have
 considered the $\BS$ as being a mixture of $b\bar{s}$ and $(bq)(\bar{s}\bar{q})$ 
states. They find that the mass of the $\BS$ state is smaller than the $BK$ 
threshold only when the four-quark componet is smaller than 30\%. For a state 
where the four-quark component is dominante (51\%) they get a mass $6.17~\GeV$.

Since the prediction for the mass of a $\BS$ scalar meson with a dominante 
four-quark component is above the $BK$ threshold  at 5774 MeV, the decay
$\BS\to BK$ will be super allowed for a $\BS$ scalar meson with a dominante 
four-quark component. As a consequence such state will be very broad, like the 
light scalars, and very difficult to be observed. 
However, a two-quark state with a mass above the $BK$ threshold could still
be narrow if the coupling at the  $B_{s0}B K$ vertex is not very large.
Therefore, in the next
section we will evaluate the coupling at the $B_{s0}B K$ vertex,
considering the $B_{s0}$ scalar meson as a  $b\bar{s}$ state.

\section{The sum rule for the  $ B_{s0}B K $  vertex with $B_{s0}$ being a 
$b\bar{s}$ state}

The coupling at the $B_{s0}B K$ vertex can be evaluated by using
the three-point function QCDSR. Here
we use the same technique developed in
previous work for the evaluation of  the couplings in the vetices
 $D^* D \pi$ \cite{nnbcs00,nnb02}, 
$D D \rho$\cite{bclnn01},  $D D J/\psi$ \cite{mnns02},  $D^* D J/\psi$ 
\cite{mnns05},  $D^* D^* \pi$ \cite{cdnn05}, 
$D^* D^* J/\psi$ \cite{bcnn05},  $D_s D^* K$, $D_s^* D K$ \cite{bccln06},  
$D D \omega$  \cite{hmm07}, $D^*D^*\rho$ \cite{bcnn08}  and $D_{sj} D K $ 
\cite{ko}.

\subsection{Sum rules for the form factors}

The three-point function associated with the $B_{s0}B K$ vertex, 
for an off-shell $B$ meson, is given by
\begin{equation}
\Gamma_{\mu}^{(B)}(p,\pli)=\int d^4x \, d^4y \;\;
e^{i\pli\cdot x} \, e^{-i(\pli-p)\cdot y}
\langle 0|T\{j_{\mu}^{K}(x) j^{B\dagger}(y) 
 j^{B_{s0}\dagger}(0)|0\rangle, 
\label{correboff} 
\end{equation}
and for an off-shell ${K}$ meson:
\begin{equation}
\Gamma_{\mu}^{({K^{0}})}(p,\pli)=\int d^4x \, 
d^4y \;\; e^{i\pli\cdot x} \, e^{-i(\pli-p)\cdot y}\;
\langle 0|T\{j^{B}(x)  j_{\mu}^{K \dagger}(y) 
 j^{B_{s0} \dagger}(0)\}|0\rangle\, .\label{correkoff} 
\end{equation}

The general expression for the vertices 
(\ref{correboff}) and (\ref{correkoff}) has two independent Lorentz structures. 
We can write each $\Gamma_{\mu}$ in terms of the invariant amplitudes associated 
with each one of these structures in the following form:

\beq
\Gamma_{\mu}(p,\pli)= F_1(p^2 , \ql , q^2)  p_{\mu}+ F_2(p^2,\ql, q^2) \pli_{\mu},
\label{gama}  
\eeq
where $q=\pli-p$.

Equations~(\ref{correboff}) and 
(\ref{correkoff}) can be calculated in two different ways: using quark 
degrees of freedom --the theoretical or OPE side-- or using hadronic 
degrees of freedom --the phenomenological side.

The phenomenological side of the vertex function, $\Gamma_{\mu}
(p,p^\prime)$,
is obtained by the consideration of $K$ and $B$ states contribution to
the matrix element in Eqs.~(\ref{correboff}) and 
(\ref{correkoff}). The coupling at the vertex $B_{s0}BK$ is defined through
the   following effective Lagrangian
\beq
\mathcal{L}_{B_{s0} B K}=
g_{B_{s0} B K} \Big ( {\bar {B_{s0}}} \bar{B} K + B_{s0} B \bar{K} \Big ), 
\label{lagran}
\eeq
from where one can deduce the matrix elements associated with the
$B_{s0} B K$ momentum dependent vertices, that can be written it in terms of the 
form factors:
\beq
\langle B_{s0}(p)|K(p') B(q)\rangle = 
g^{(B)}_{B_{s0}BK}(q^2),
\label{ffb}
\eeq
and
\beq
\langle B_{s0}(p)|B(p') K(q)\rangle = 
g^{(K)}_{B_{s0}BK}(q^2).
\label{ffk}
\eeq

The meson decay constants, $f_{B_{s0}}$, $f_{B}$ and $f_{K}$, are
defined by the following  matrix elements:
\beq
\langle 0|j^{{B}_{s0}}|{B_{s0}(p)}\rangle= m_{B_{s0}} f_{B_{s0}} ;
\label{fBs0}
\eeq
\beq
\langle 0|j^{B}|{B(p)}\rangle= \frac{m^2_{B}}{m_b} f_{B}
\label{fB}
\eeq
and
\beq
\langle 0|j_{\nu}^{K}|{K(p)}\rangle= i f_K p_{\mu} ,
\label{fK}
\eeq 
Saturating Eqs.~(\ref{correboff}) and  (\ref{correkoff}) with 
$B $ and $K$ states and using Eqs.~(\ref{ffb}), (\ref{ffk}), (\ref{fBs0}),
(\ref{fB}) and (\ref{fK}) we arrive at
\beq
\Gamma_{\mu  }^{(B)phen}(p,\pli)= g^{(B)}_{B_{s0} B K}(q^2) 
 \frac{f_{B_{s0}} f_{K} f_{B} \frac{m^2_{B}}{m_b} m_{B_{s0}} m_{K}}
{(p^2-m^2_{B_{s0}})(q^2-m^2_{B})({\pli}^2 -m^2_{K})} \pli_{\mu},
\label{phenboff}
\eeq
when the $B$ is off-shell, with a similar expression for the
$K$ off-shell: 
\beq
\Gamma_{\mu  }^{(K)phen}(p,\pli)= g^{(K)}_{B_{s0} B K}(q^2) \frac{f_{B_{s0}} 
f_{B} f_K \frac{m^2_{B}}{m_b} m_{B_{s0}} m_{K}}
{(p^2-m^2_{B_{s0}})(q^2-m^2_{K})({\pli}^2 -m^2_{B})} q_{\mu}.
\label{phenkoff}
\eeq

In the OPE or theoretical side  the currents appearing in Eqs.~(\ref{correboff}) 
and (\ref{correkoff}) can be written 
in terms of the quark field operators in the following form: 
\beq
j_{\mu}^{K}(x) = \bar s(x) \gamma_{\mu} \gamma_{5} u(x);
\label{cok}
\eeq 

\beq
j^{B}(x) = i \bar b(x) \gamma_{5} u(x) 
\label{coB}
\eeq
and 
\beq
j^{B_{s0}}(x) = \bar s(x)  b(x), 
\label{coBs0}
\eeq
Each one of these currents has the same quantum numbers of the associated
meson. 
 
For each one of the invariant amplitudes appearing in Eq.(\ref{gama}), we 
can write a double dispersion relation over the virtualities $p^2$ and 
${\pli}^2$, holding $Q^2= -q^2$ fixed:
\begin{equation}
F_i(p^2,{\pli}^2,Q^2)=-\frac{1}{4\pi^2}\int_{s_{min}}^\infty ds
\int_{u_{min}}^\infty du \:\frac{\rho_i(s,u,Q^2)}{(s-p^2)(u-{\pli}^2)}\;,
\;\;\;\;\;\;i=1,2 \label{dis}
\end{equation}
where $\rho_i(s,u,Q^2)$ equals the double discontinuity of the amplitude
$F_i(p^2,{\pli}^2,Q^2)$, calculated using the Cutkosky's rules. 

We can work with any structure appearing in
Eq.(\ref{gama}). However, since in Eq.~(\ref{phenboff}) only the $\pli_\mu$ 
structure
appears we choose to work with the $ \pli $  structure. In order to
reduce the influence of higher resonances and the pole-continuum transition
constributions  we perform a  double Borel transform 
in both variables
$P^2=-p^2\rightarrow M^2$ and ${P^\prime}^2=-{\pli}^2\rightarrow \mli$ and 
equate the two representations
described above. We get the following sum rules:
\beqa
& & \frac{m_{B_{s0}} m^2_{B}}{m_b} f_{K} f_{B}f_{B_{s0}} g^{(B)}_{B_{s0}
 BK}(Q^2) 
e^{ -m^2_{B_{s0}}/M^{2} } e^{ - m^2_{K}/M^{'2} }  =  
(Q^2 + m^2_{B})  \,\,  \bigg[ m_b<\bar{s}s> e^{-m^2_b/M^{2}} \nonumber \\
& & - \frac{1}{4 \pi^2} \int_{m^2_b}^{s_0} 
ds  \int_0^{u_{max}} \, 
d u \exp(-s/M^2) \exp(-u/M^{'2}) 
f(s,t,u) \theta(u_0 -u) \bigg]
\label{boff}
\eeqa
for an off-shell $B$, and
\beqa
& & \frac{m_{B_{s0}} m^2_{B}}{m_b} f_{K} f_{B}f_{B_{s0}} g^{(K)}_{B_{s0}
 BK}(Q^2) 
e^{ -m^2_{B_{s0}}/M^{2} } e^{ - m^2_{B}/M^{'2} }  
\nn\\
&=&  -
{Q^2 + m^2_{K}\over4 \pi^2} \int_{m^2_b}^{s_0} 
ds  \int_{u_{min}}^{u_0} \, 
d u e^{-s/M^2} e^{-u/M^{'2}} 
g(s,t,u)  
\label{koff}
\eeqa
for an off-shell $K$.

In Eqs.~(\ref{boff}) and (\ref{koff}), $t=-Q^2$, 
\beqa
f(s,t,u)&=&{3\over2[\lambda(s,u,t)]^{1/2}}\left(m_b^2+2m_bm_s-s+\right.
\nn\\
&+&\left.{(2m_b^2+2m_bm_s-s-t+u)(m_b^2(s-t+u)+s(t+u-s))\over\lambda(s,u,t)}
\right),
\eeqa
\beqa
g(s,t,u)&=&{3\over[\lambda(s,u,t)]^{3/2}}\left[m_b^4(s-t+3u)+
u\left(m_bm_s(s+t-u)\right.\right.
\nn\\
&+&\left.\left.s(-s+t+u)\right)+m_b^2\left(-2u(s-t+u)+m_bm_s(s-t+3u)\right)
\right].
\eeqa
$\lambda(s,u,t)=s^2+u^2+t^2-2su-2st-2tu$,  $u_{min}=m_b^2-{m_b^2t\over s- 
m_b^2}$ and $u_{max}=s+t-m_b^2-{st\over m_b^2}$.

Since we are dealing with heavy quarks, 
we  expect  the perturbative contribution to dominate  the 
OPE. For this reason, we do not 
include the gluon and quark-gluon condensates in the present work.

In this work we use the following relations between the Borel masses $M^2$ and 
$M'^2$: $\frac{M^2}{M'^2} = \frac{m^2_{B_{s0}}-m_b^2}{0.64 \GeV^2}$
for a  $B$ off-shell  and $\frac{M^2}{M'^2} = \frac{m^2_{B_{s0}}}{m^2_{B}}$ for 
a $K$ off-shell.

\subsection{Results for the form factors}

\begin{table}[h]
\begin{tabular}{|c|c|c|c|c|c|c|c|} \hline
$m_b (\GeV)$ & $m_{B_{s0}} (\GeV)$ & $m_{B} (\GeV)$ & $m_{K} (\GeV)$ & 
$f_{B_{s0}} (\GeV)$ &$f_{B} (\GeV)$ & $f_{K} (\GeV)$ & 
$\langle \bar q q \rangle (\GeV)^3$  \\ \hline 
4.7& 5.70 &5.28 &0.49 & 0.24&0.17& 0.16 &$(-0.23)^3$ \\ \hline
\end{tabular}
\caption{Parameters used in the calculation.}
\label{tableparam}
\end{table}

Table \ref{tableparam} shows the values of the parameters used in the present 
calculation. We take $f_{B_{s0}}$ and $m_{B_{s0}}$ and $m_b$ from 
ref.~\cite{wang}, where
a QCDSR calculation is used to study the $\BS$ considered as a $b\bar{s}$ scalar 
meson. The continuum thresholds are given by 
$s_0=(m_{\BS} + \Delta_s)^2$ and $u_0=(m+\Delta_u)^2$, where $m$ 
is the kaon  mass, for a  $B$ off-shell  and the $B$ meson mass, for 
a $K$ off-shell.

\begin{figure}[ht] 
\begin{center}
\centerline{\epsfig{figure=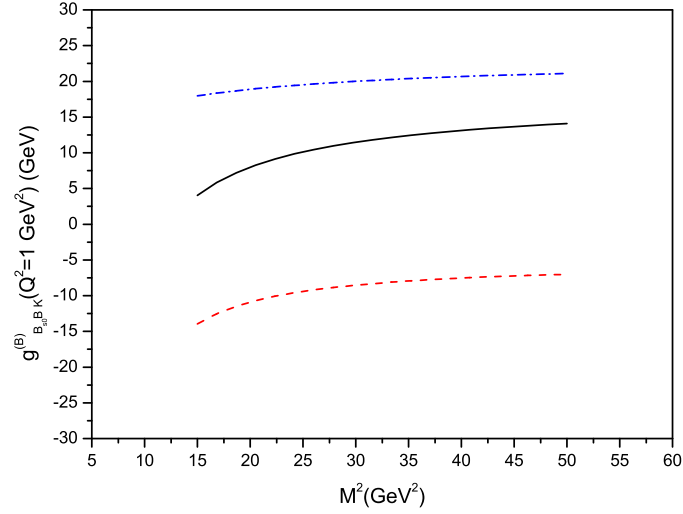,height=80mm}}
\caption{$g^{(B)}_{B_{s0} B K}(Q^2=1.0 GeV^2)$ as a function of the Borel mass 
$M^2$. The dot-dashed dashed and solid lines 
correspond to the perturbative, quark condensate and total contributions 
respectively.}
\label{fig3}
\end{center}
\end{figure}

Using  $\Delta_s=\Delta_u = 0.5 \GeV $ for the continuum thresholds
and fixing $Q^2=1 \GeV^2$, we found a good stability of the 
sum rule for $g_{B_{s0} B K}^{(B)}$ for $M^2\geq 20\GeV^2$, as can be seen in 
Fig.~\ref{fig3}.

%===========================================================================

Within this interval 
we need to choose the best value of the  Borel mass to 
study the $Q^2$ dependence of  the  form factor. 
It is well known in QCDSR that for  small values of the Borel 
variable, $M^2$, the sum rule is dominated by the pole. However, the 
convergence of the OPE always get better for large values of $M^2$. On the other 
hand, for very large values of $M^2$  the OPE convergence is perfect but the 
sum rule is dominated by the continuum. The best value of the Borel mass is the 
one for which one has a good OPE convergence and the pole contribution is bigger 
than the continuum contribution. In this case the both criteria are  
reasonably satisfied for $M^2 \approx 35~\GeV^2 \approx m^2_{B_s0}$.

In the case of $g_{B_{s0} B K}^{(K)}(Q^2)$, doing a similar analysis described 
above, we also fix  $M^2 \approx 35~\GeV^2 \approx m^2_{B_s0}$.

\begin{figure}[ht] 
\epsfig{file=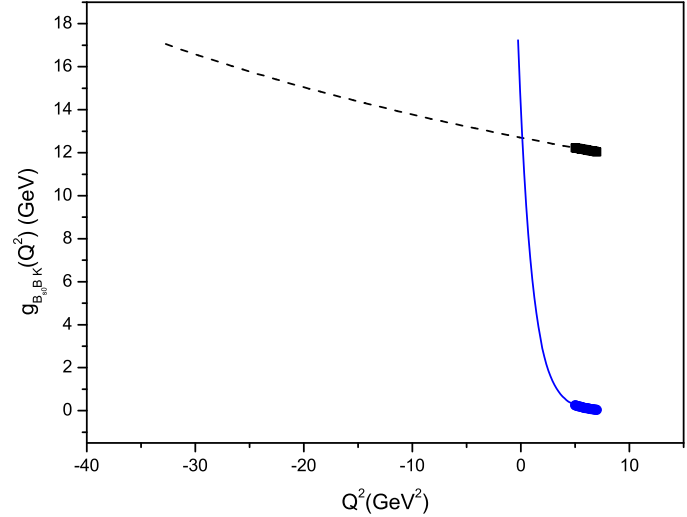}
\caption{$g^{(B)}_{B_{s0} B K}$ (squares) and 
$g^{(K)}_{B_{s0} B K}$ (circles) QCDSR form factors as a function of
$Q^2$. The dashed and solid lines 
correspond to the monopolar and exponetial parametrizations respectively.}
\label{fig4}
%\end{center}
\end{figure}

Having determined $M^2$ we show, in Fig.~\ref{fig4}, the $Q^2$ dependence of the 
form factors. The squares correspond to the 
$g_{B_{s0} B K }^{(B)}(Q^2)$ form factor in the  interval where the sum rule is 
valid. 
The circles are the results of the sum rule for the $g_{B_{s0} B K}^{(K)}(Q^2)$ 
form factor. 

\vspace{0.5cm}
In the case of an off-shell $ B $ meson, our numerical results can be 
fitted by the following monopolar parametrization (shown by the dashed line in
Fig.~\ref{fig4}):
\begin{equation}
g_{B_{s0} B K}^{(B)}(Q^2)= \frac{1,629.12\GeV^3} {Q^2 + 128.25\GeV^2} \;.
\label{monoboff}
\end{equation}

The coupling constant is defined as the value of the 
form factor at $Q^2= -m^2_{m}$, where $m_{m}$ is the mass of the off-shell meson. 
Therefore, using $Q^2=-m_{B}^2$ in Eq~(\ref{monoboff}), the resulting coupling 
constant is:
\begin{equation}
g_{B_{s0} B K}= 17.01~\GeV \;.  
\label{couplingboff}
\end{equation}

For an off-shell $K$ meson, our sum rule results  can also be 
fitted by an exponential parametrization, which is represented by the
solid line in Fig.~\ref{fig4}:  
\begin{equation}
g_{B_{s0} B K}^{(K)}(Q^2)= 14.14\GeV e^{-Q^2/1.25\GeV^2}\;.
\label{expkoff}
\end{equation}
Using $Q^2=-m_{K}^2$ in Eq~(\ref{expkoff}) we get:
\begin{equation}
g_{B_{s0} B K}= 17.23~\GeV,
\label{couplingkoff}
\end{equation}
in a good agreement with the result of Eq.(\ref{couplingboff}).

\begin{figure}[ht] 
\epsfig{file=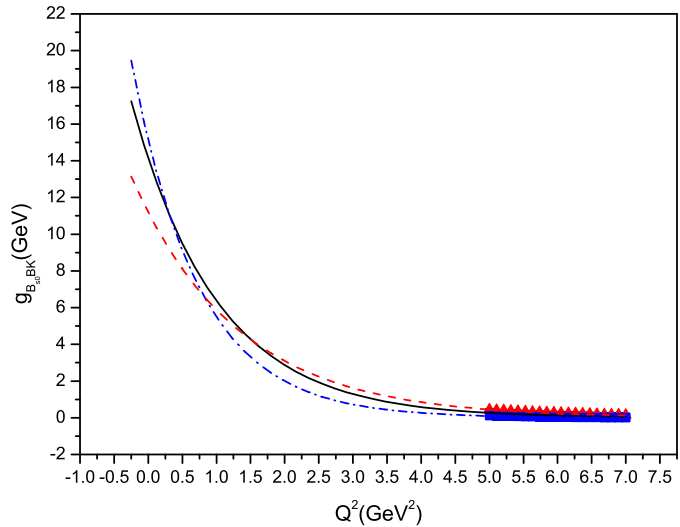} 
\caption{Dependence of the form factor on the continuum threshold for
the $K$ off-shell case. The solid curve  corresponds to 
$\Delta_{s,u} = 0.5\GeV$, the dashed one  to $\Delta_{s,u}= 0.6\GeV$
and the dotted curve to $\Delta_{s,u}= 0.4\GeV$.}
\label{fig5}
\end{figure}

In order to study the dependence of our results with the continuum
threshold, we vary $\Delta_{s,u}$ between 
$0.4\GeV\le \Delta_{s,u}\le 0.6\GeV$ in the parametrization
corresponding to the case of an  off-shell  $K$. As can be seen in 
Fig.~\ref{fig5}, this procedure gives us an uncertainty interval of 
$ 13.13\GeV \le g_{B_{s0} B K} \le 19.46\GeV$ for the coupling constant.

We see that in the two cases considered here, off-shell $B $ or $K $,
we get compatible results for the coupling constant, evaluated using the QCDSR 
approach. Considering the uncertainties in the continuum
thresholds we obtain:
\begin{equation}
g_{B_{s0} B K}= (16.29 \pm 3.16)~\GeV\;. \label{finalcoupling}
\end{equation}
in a very good agreement with the coupling evaluated in ref.~\cite{wang2}
using light-cone QCDSR: $g_{B_{s0} B K}= (19.6 \pm 5.7)~\GeV$.

In ref.~\cite{gscpz} the authors use a heavy-light chiral lagrangian and find
the $B_{s0}$ to be a $BK$ bound state with the strong coupling
$g_{B_{s0} B K}= 23.442~\GeV$. This coupling is bigger than our result in 
Eq.~(\ref{finalcoupling}), which is compatible with the expectation that a
multiquark system would decay easily into its constituents. 

The decay width for the decay $\BS\to BK$ is given in terms of the coupling 
constant $g_{B_{s0} B K}$ through:
\beq
\Gamma(\BS\to BK)={1\over16\pi m_{\BS}^3}g_{\BS BK}^2\sqrt{\lambda(m_{\BS}^2,
m_B^2,m_K^2)},
\enq

Considering the result in Eq.~(\ref{finalcoupling}) and using two different 
values for $m_{\BS}$, we give in Table \ref{tabga} the predictions for
the  $\BS\to BK$ decay width. We also include in this Table, the decay width
obtained with the result from ref.~\cite{gscpz}.

\begin{table}[h]
\begin{tabular}{|c|c|c|}  \hline
$m_{\BS}~(\GeV)$ & $\Gamma~(\GeV)$ (this work) & $\Gamma~(\GeV)$ 
(ref.~\cite{gscpz})  \\
\hline
 5.775 & $10\pm4$ & 20\\
\hline
 5.8 & $50\pm15$ & 100\\
\hline
\end{tabular}
\caption{The  $\BS\to BK$ decay width.}
\label{tabga}
\end{table}

\section{Conclusions}

We have used the QCDSR to study the two-point function for the $\BS$ scalar 
meson, considered as a tetraquark state in a diquark-antidiquark configuration.
 The mass obtained for
the $(bq)(\bar{q}\bar{s})$ scalar state: $m_{\BS}=(5.85\pm0.13)~\GeV$, is in
agreement with  other predictions using different structures. Therefore, if the 
$\BS$ scalar meson is observed with a mass around 5.7 -- 5.8 GeV, only this 
information will not be enough to discriminate its structure. However, the
width of the state can also be used to help in this task. For this
purpose we have also considered the QCDSR three-point function for the vertex 
$\BS BK$ to evaluated the $\BS BK$ coupling constant,
considering the $\BS$ scalar meson  to be a  normal $b\bar{s}$ state.
With this configuration we find the coupling constant at the $\BS BK$
vertex to be $g_{B_{s0} B K}= (16.29 \pm 3.16)~\GeV$, in a very good agreement
with calculation using light-cone QCDSR. 

In Table~\ref{tabga} we have presented the predictions for the $\BS\to BK$
decay width, using two different values for $m_{\BS}$, and the couplings
obtained considering the $\BS$ as a $b\bar{s}$ state and a  $BK$ bound state.
As expected the width obtained in the case that $\BS$ state is a multiquark
state is much bigger than the width obtained for a $b\bar{s}$ state with the 
same mass. Therefore,
with the knowledge of the decay width and the mass of the $\BS$ it will be 
possible to discriminate between possible structures for this state.

\acknowledgments
The authors would like to thank Fernando S. Navarra for usefull conversations. 
This work has been supported by CNPq and FAPESP.

\end{document}